\documentstyle[pre,preprint,aps]{revtex}

\begin{document}
\draft
\title{Coin Tossing as a Billiard Problem}
\author{N. L. Balazs, Rupak Chatterjee, and A. D. Jackson}
\address{
  Department of Physics, State University of New York at Stony Brook,
  Stony Brook, New York 11794-3800, USA}
\date{\today}
\maketitle

\begin{abstract}
We demonstrate that the free motion of any two-dimensional
rigid body colliding elastically with two parallel, flat walls is
equivalent to a billiard system. Using this equivalence, we analyze
the integrable and chaotic properties of this new class of billiards.
This provides a demonstration that coin tossing,
the prototypical example of an independent random process, is
a completely chaotic (Bernoulli) problem.  The related question of
which billiard geometries can be represented as rigid
body systems is examined.
\end{abstract}


\section{Introduction}
\label{Intro}
Classical dynamics can describe chaotic motion.  It is generally accepted that
truly chaotic systems are isomorphic to Bernoulli systems.  This means that any
chaotic solution is equivalent in a well-defined sense to a stochastic one
generated by an independent random process.  The prototype of such a process is
coin tossing described by repeated random trials with only two possible
outcomes for each trial with probabilities which do not change during the
trials.  These trials are called Bernoulli trials, hence the name.  However,
the tossing of a coin can also be viewed as a simple dynamical system since it
is merely the motion of a rigid body subjected to simple forces and boundary
conditions.  It is the legitimate to ask whether coin tossing is {\em de
facto\/} a Bernoulli system when considered as a dynamical process.  The
experience of gamblers through the ages would suggest that it is.  The aim of
the present paper is to provide a more mathematical demonstration.

To be specific, we shall consider a rigid body in two dimensions which moves in
the absence of forces and makes elastic collisions with two infinitely massive
parallel walls.  The special case in which this rigid body
is simply a stick of zero width will serve to model coin tossing.  We
shall demonstrate that this system is equivalent to that of a point particle
colliding with a suitably curved wall. The key feature to this
equivalence is the fact that elastic collisions by the rigid body
necessarily imply {\em specular\/} reflections for the related
point particle, {\em i.e.}, the behaviour of a traditional billiard.  For the
special case of tossing a coin between plane walls, the equivalent billiard
configuration is that of a particle bouncing
between two convex sinusoidal boundary walls.  It has been shown in \cite{gal}
that billiards with strictly convex walls are of the Bernoulli type
indicating that coin tossing, when considered as a dynamical system,
is truly a completely random process. However, for billiards arising from
coin tossing, the correlation between the initial
orientation of the coin and its orientation on subsequent bounces dies out
with an exponential envelope. This then provides a characteristic
decorrelation time after which the process is random.
A fair coin toss must necessarily allow the
coin to bounce repeatedly on the floor and the ceiling.  Evidently, a
simple toss and catch process, the colloquial example of a random process,
does not suffice.

We shall see that every such rigid body system is equivalent to a unique
billiard problem.  The generality of this equivalence is of some interest.
For some time, billiard problems have been known to provide
convenient and illustrative examples of the general properties
of Hamiltonian dynamical systems \cite{sin,bun,ben}.
The quantum properties of billiards have also been studied (see,
for instance \cite{boh}).
Physical realizations of billiard systems have been found in microwave
cavities as well as in electronic nanometric semiconductor devices
and used to investigate how the transition from regular
to chaotic motion influences the wave properties of these systems
\cite{dor,jal,mar}.  Yet, many of the billiard geometries studied
are somewhat arbitrary and do not represent real physical systems.  The present
results offer a broad class of physically realizable dynamical systems
exhibiting all types of motion; integrable, near integrable (KAM), and chaotic
motion of increasing randomness ($K$-flows, $C$-flows, and Bernoulli flows).

The organization of the paper is as follows. In section \ref{trans}, we
consider the specific problem of the stick (or coin).  The
general transformations relating the rigid body problem to the
billiard system are derived, and the condition for specular reflection
is proved. Section \ref{elps} is deals with some of the properties
associated with elliptical rigid bodies.  In particular, we consider how
the onset of chaos depends on the eccentricity
of the ellipse and the separation between the walls.  In section \ref{shape},
we consider the inverse problem of deriving body shapes from given wall
shapes.  Finally, a number of conclusions are
drawn in section \ref{endup} along with some suggestions for certain
interesting shapes and possible directions for a quantum mechanical
analysis of this problem.

\section{Transformation Equations}
\label{trans}

\subsection{Sticks}
\label{sub:sticks}

Let us first consider the simple case of a stick of length $2L$,
mass $M$, and radius of gyration $\kappa $, bouncing freely between two
flat walls separated by a distance $H$. (For simplicity, we assume that the
center of mass is at the mid-point of the stick.)  The position of the stick
is characterized by $y$, the height of the center of mass above the
lower wall, and $\theta$, the angle of rotation of the stick from
the vertical. If we introduce the dimensionless height $\eta = y/\kappa$,
the (scaled) energy of the stick is given by
\begin{equation}
E = \frac{M}{2} ( \dot{\theta ^{2}} + \dot{\eta ^{2}}).
\end{equation}
This looks formally like the energy of a free point particle
moving in an Euclidean plane parameterized by the dimensionless
coordinates $x_1 = \theta$ and $x_2 = \eta$. At the collision
of the stick with the wall, it is easy to see that
\begin{eqnarray}
(M\kappa )\delta \dot{\eta } & = & f_n \delta t, \nonumber\\
(M \kappa ^{2})\delta \dot{\theta } & = & L\sin(\theta )f_n \delta t,
\end{eqnarray}
where $f_n$ is the normal force at the wall. This gives a linear relation
between the change in velocities at the collision point,
\begin{equation}
\delta \dot{\theta } = \frac{L}{R} \sin(\theta ) \delta \dot{\eta }.
\end{equation}
Now, the boundary for the motion of our point particle is given
by the minimum value of $\eta $ for a given angular orientation.
{}From conservation of energy we find
(using (1) and (3)) that
\begin{equation}
\delta \dot{\eta } = \frac{-2(\dot{\eta } + (L/\kappa)
\dot{\theta } \sin(\theta ))}{1 + (L/\kappa)^{2}
\sin(\theta )^{2}},
\end{equation}
resulting in $\eta _{\rm min} = (L/\kappa )|\cos(\theta )|$.
Therefore, the bottom and top boundary walls ($b ( \theta )$ and
$t( \theta )$, respectively) corresponding to the billiard problem
for the stick are
\begin{eqnarray}
b(\theta ) & = & \frac{L}{\kappa }|\cos(\theta )|, \nonumber\\
t(\theta ) & = & H - \frac{L}{\kappa }|\cos(\theta )|.
\end{eqnarray}

Finally, consider the infinitesimal change in the velocity vector
of our point particle after a collision,
\begin{equation}
\delta \vec{v} = (\vec{T} \cdot \delta \vec{v}) \vec{T}
+ (\vec{N} \cdot \delta \vec{v}) \vec{N}.
\end{equation}
Here, $\vec{T}$ and $\vec{N}$ are the orthonormal tangent and normal
vectors at the billiard boundary.  The tangent vector along the
lower wall is
\begin{equation}
\vec{T} = (1,-\frac{L}{\kappa }\sin(\theta )).
\end{equation}
Using (3), we have
\begin{equation}
\vec{T} \cdot \delta \vec{v} = \vec{T} \cdot (\delta \dot{\theta },
\delta \dot{\eta }) = 0,
\end{equation}
and consequently, the tangential component of $\delta \vec{v}$
vanishes. Since the magnitude of $\vec{v}$ is preserved by energy
conservation, the above result indicates that upon collision, the
normal component of $\delta \vec{v}$ simply changes sign. Thus,
the point particle in our equivalent undergoes {\em specular\/} reflections
at the boundary walls. Hence, the motion of a rigid stick is equivalent
to a billiard problem with properly chosen boundaries. (Since
the horizontal $\theta $-axis is clearly periodic, one may
consider the configuration space to be the surface of a cylinder
with `undulating' top and bottom edges.)

\subsection{Arbitrary Shapes and Mass Distributions}
\label{sub:arbit}

It is a simple matter to generalize the above to the case of
rigid bodies of arbitrary shape. We restrict our discussion
to objects possessing at least one symmetry axis in the plane. (The
procedure for asymmetric objects is identical except that the upper
boundary is shifted by $\pi$ with respect to the lower.)  Consider a
closed plane curve $\cal C$.  If $\cal C$ is not everywhere convex, we
replace $\cal C$ with its convex envelope since concave segments
cannot come into contact with the boundary walls. As before,
the inertial properties of the object will be expressed in terms
of the total mass $M$ and the radius of gyration $\kappa $. The
height of the center of mass of the object above the wall
will be denoted by $\eta $ (scaled by the gyration factor $\kappa $).
The curve ${\cal C}$ can be specified completely by its radius, $R( \alpha )$,
as a function of the angle $\alpha$ from some body-fixed axis ${\hat a}$.
\begin{equation}
\vec{R}(\alpha) = (R(\alpha )\cos(\alpha ),\ R(\alpha )\sin(\alpha ))
\end{equation}
It is elementary to construct the tangent to ${\cal C}$, defined as
\begin{equation}
{\vec T}( \alpha ) = d{\vec R} / d\alpha \ \ ,
\end{equation}
and the associated normal vector ${\vec N}( \alpha )$.  When the point of
contact between the body and the wall is at $\alpha_c$, the perpendicular
distance between the center of mass and the wall is simply
\begin{equation}
b( \alpha_c ) = \frac{1}{\kappa} \ \frac{{\vec N} \cdot {\vec R}}{( {\vec N}
\cdot {\vec N} )^{1/2}} = \frac{1}{\kappa} \ \frac{R^2}{\sqrt{R^2 + R'^2}} \ \
{}.
\end{equation}
To obtain the desired wall shape, we must determine that orientation of the
rigid body, $\theta$, which makes $\alpha_c$ the point of contact.  (For
convenience, we measure $\theta$, clockwise, from the normal to the bottom
wall to the body-fixed axis ${\vec a}$.)
Obviously, the condition of tangency
is that ${\vec N}( \alpha_c )$ is parallel to the wall normal:
\begin{equation}
\frac{R'(\alpha _c)}{R(\alpha _c)} = \tan(\alpha _c + \theta ),
\end{equation}
which then produces the lower boundary $b(\theta ) =
\eta (\alpha _c (\theta))$ with the upper wall being given by
$t(\theta ) = H - b(\theta )$ (or $t(\theta )
= H - b(\theta ) + \pi$ for asymmetric objects).

The equations describing the actual collision have a form similar
to equations (2),
\begin{eqnarray}
(M\kappa )\delta \dot{\eta } & = & f_n \delta t, \nonumber\\
(M \kappa ^{2})\delta \dot{\theta } & = &
R(\alpha _c)\sin(\alpha _c + \theta )f_n \delta t,
\end{eqnarray}
with the energy again being given by (1). Since equation (13)
follows from (2) with the substitutions
\begin{eqnarray}
L & \rightarrow & R(\alpha_c) \ \ , \nonumber\\
\theta & \rightarrow & \theta + \alpha_c \ \ ,
\end{eqnarray}
it is clear that the proof for specular reflection is identical
to that given for the stick.  Thus, for {\em every\/} shape of the rigid body,
there is an equivalent billiard problem.

As an example of the above procedure, consider an elliptical
disk with a semi-major axis of one and a semi-minor axis
of $1/(\sqrt{1 + \epsilon})$. Imagine that the center of mass
is at the geometric center of the ellipse. Relative to this
point, the ellipse is parameterized as
\begin{equation}
R(\alpha ) = \frac{1}{(1 + \epsilon \sin ^{2}(\alpha ))^{1/2}}.
\end{equation}
Equation (12) leads to
\begin{equation}
\cos(\alpha _c + \theta ) = \frac{(1 + \epsilon \sin ^{2}(\alpha _c ))}
{(1 + (\epsilon ^{2} + 2\epsilon ) \sin ^{2}(\alpha _c ))^{1/2}}
\end{equation}
and
\begin{equation}
\sin ^{2}(\alpha _c) = \frac{\sin ^{2}(\theta)}{(1
+ (\epsilon ^{2} + 2\epsilon ) \cos ^{2}(\theta ))}.
\end{equation}
Substituting (15), (16), and (17) into (11) and (12) results in the following
billiard boundary
\begin{equation}
b(\theta ) = \left( \frac{1}{\kappa } \right) \sqrt{\frac{
(1 + \epsilon \cos ^{2}(\theta ))}{1 + \epsilon}}.
\end{equation}
In the limit of $\epsilon \rightarrow 0$, $b(\theta )
\rightarrow 1/\kappa $ which is the obvious result for a circle
of radius one. In the other limit $\epsilon \rightarrow
\infty $, the ellipse degenerates into a stick and we recover
the previous result of $b(\theta ) = (1/\kappa ) |\cos(\theta )|$.

\section{Elliptic Discs}
\label{elps}

{\em KAM motion\/}:  We now focus on the
special case in which the rigid body is an ellipse
so that the equivalent billiard problem has the wall described by
equation (18) of the preceding section.  A Poincare section map for the
elliptic system with $\epsilon = 0.003$ and $H = 3$ is shown in Fig.2. This
is a phase portrait
of the contact point $\theta _c $ versus the angle of incidence
$\phi _c $ with respect to the local normal at $\theta _c $.
The regular and chaotic regions of this ``mixed'' system are
clearly visible. For such a small value of the eccentricity
$\epsilon $, the boundary can be expanded as
\begin{equation}
b(\theta ) \simeq (1 - (\epsilon /2) \sin ^{2}(\theta )).
\end{equation}
This indicates that even a small variation from the integrable
billiard generated by a circle can produce a dense population of
stochastic trajectories. The corresponding phase space can be
described using the KAM theorem \cite{arn}. For instance, Fig.3
is a phase portrait for $\epsilon = 0.55$ and $H = 1.9$. This
section map contains both hyperbolic and elliptical regions
of phase space. The inner circles correspond to KAM tori
of irrational winding number frequencies.
The island chains encircling these tori surround the
elliptic fixed points of the motion. As is well known, this
structure is self-similar and persists at smaller length scales
{\em ad infinitum\/}.

The transition from a mixed geometry to a chaotic one occurs
suddenly in these rigid body problems.  As long as the motion is purely
chaotic,
the actual value of $H$ plays no explicit role.  When the
motion contains both hyperbolic and elliptic regions, the relative
size of these different regions is dictated by the height $H$ with the
elliptic regions disappearing as $H$ and/or $\epsilon$ become larger.
It is somewhat delicate to
make a numerical determination of the curve which separates the KAM and
chaotic regions.  This is particularly true when $H$ is large and
$\epsilon$ is small.  The approximate boundary curve for the elliptic
problem is
\begin{equation}
\epsilon \simeq \frac{1}{H} + \frac{17}{H^2} \ \ .
\end{equation}
This expression reproduces the results of simulations to about 1\% which is
probably the limit of their accuracy.   In the limit as $\epsilon \rightarrow
0$ and the object approaches a
circle, the wall function approaches a constant plus a small sinusoidal
oscillation.  Given any $\epsilon$, equation (20) shows us that it is always
possible to pick $H$ sufficiently large that K-type motion results.

{\em Application to coins\/}:  The case of a very eccentric ellipse
can be viewed as a model of a coin tossing process. Regarding this ellipse
as a coin observed edgewise, we place an oriented arrow along its
length. With this fixed orientation, we assign the condition
``heads'' when the arrow points to the right and ``tails''
when the arrow points to the left. The motion of the coin consists
of the one-dimensional vertical motion of the center of mass and a
superimposed rotation. This leads to a strictly convex
billiard problem $b(\theta ) = |\cos (\theta )|$ which has been shown
\cite{gal} to be isomorphic to a Bernoulli shift.  Surprisingly, this
includes the case where the distance between the walls is less than the length
of the stick.  (In our coin analogy, this means that the room is not high
enough to permit the coin to flip from heads to tails.)  In this case, the
full angular region is not accessible.  Nevertheless, the motion is chaotic.
This is consistent with equation (20) which indicates that, for any $H$, it
is always possible to pick $\epsilon$ sufficiently large that K-type motion
results.

In spite of its common use as a random process, the traditional tossing of a
coin (without bounces) is nothing of the sort.  As we have shown, the
present coin toss (including collisions with a wall) does result in K-type
motion.  In this regard, it is interesting to consider the probability, $P(n)$,
that, if the coin shows a ``head'' at a given collision with the bottom wall,
it will also
show a ``head'' $n$ collisions later.  We expect $P(n) \rightarrow 1/2$ for
an independent random process.  We have studied $P(n)$ numerically as a
function of $H$ for a very eccentric ellipse ($\epsilon = 99$).
It is readily seen that $P(n)$ approaches the asymptotic value of $1/2$
exponentially.
\begin{equation}
P(n) - 1/2 \sim \exp{(-n / \tau (H) )}
\end{equation}
When the distance between the walls is equal to the diameter of the coin, the
correlation time $\tau$ is evidently infinite.  As the wall separation is
increased slightly, the correlation time is approximately inversely
proportional to the difference between $H$ and the diameter
of the coin.  For example, when $H$ is 10\% larger than the diameter of
the coin, $20$ bounces are sufficient to reduce the
non-asymptotic part of $P(n)$ to $0.01$.  When $H$ is 20\% larger than the
diameter, $12$ bounces suffice.  It requires exceptionally good statistics to
follow the non-asymptotic part of $P(n)$ when $H$ becomes appreciable.  The
``house rules'' at casinos inevitably require that dice be bounced from a wall.
The present results suggest that this bounce plays a significant role in
randomizing the outcome and is not merely a time-honored tradition.

\section{From wall shapes to body shapes}
\label{shape}

We have demonstrated that every problem involving elastic collisions between a
rigid body and parallel walls is equivalent to a billiard with a suitable
(and unique) choice of the wall function $b( \theta )$.  Here, we wish to
address the inverse problem.  For convenience, we shall restrict our attention
to periodic walls which are everywhere differentiable.  (This is not to deny
that rigid bodies with sharp corners are also of interest.)  There are
several questions of interest:  How does one
determine $R( \alpha )$ given $b( \theta )$?  Does every periodic $b( \theta )$
correspond to a realizable rigid body and, if so, is its shape unique?  The
general technique for going from $b( \theta )$ to $R( \alpha )$ is easy to
state but can be challenging to implement.  Formally, we use equation (12)
to eliminate $\theta$ from equation (11).  This leads to a non-linear but
first-order differential equation for $R( \alpha )$.  Two observations are
useful.  First, the extrema in $b$ (as a
function of $\theta$) coincide with the extrema of $R$ (as a function of
$\alpha$) with $\alpha$ and $\theta$ related by equation (12).  Second, at such
extrema, $\alpha = \theta$ and $R( \alpha ) = b( \theta )$.  While it may be
difficult to find analytic solutions to the resulting equation, numerical
solutions are accessible.  It is not always easy to find them.  For example,
imagine that the wall shape is given by equation (17).  The procedure stated
leads to a quadratic equation for $R'(\alpha)$ which is readily solved to
yield
\begin{equation}
\frac{R'}{R} = \frac{-\epsilon \sin \alpha \cos \alpha \pm
\sqrt{-1-\epsilon+(1+\epsilon)(1+\epsilon \sin^2 \alpha)R^2}}
{1+\epsilon \sin^2 \alpha}
\end{equation}
which is consistent with equation (15).  Note, however, this consistency is
realized by two degenerate solutions which renders the numerical problem
delicate.

As we shall see, every periodic $b( \theta )$ can be regarded as an equivalent
rigid body problem.  According to a well-known theorem, $b( \theta )$ must
have at least four extrema to represent a physical wall shape.  Every
non-constant, periodic function must have at least one maximum and one
minimum.  Thus, the four-extremum requirement can always be met by mapping
$\theta \rightarrow \theta / 2$.  Before proceeding, we note a fundamental
ambiguity in the inverse problem.  If one is given only the equivalent
billiard problem, the distinction between $b( \theta )$ and $H$, the separation
between the parallel walls, cannot be made uniquely.  Thus,
$b( \theta )$ can only be determined up to some constant shift $b_0$.
Due to the non-linear nature of the inverse problem, the corresponding
$R( \alpha )$ generally depends on the value of $b_0$ assumed.  Hence, the
equivalent billiard does not determine the rigid body uniquely.  Moreover,
there will generally be choices of $b_0$ for which no physically realizable
$R( \alpha )$ exists.  (The specific example of elliptic discs is considered
in the Appendix where we show that no physical solution exists when $b_0$
is less than some critical value.)

While the freedom of choosing $b_0$ and the associated multiplicity of body
shapes which arise from its exploitation may seem a unfortunate, there is
a silver lining.  This freedom allows us to see that there are always
(infinitely many) rigid body shapes corresponding to any
choice of $b( \theta )$.  Proof of existence is compensation for loss of
uniqueness.  Consider the elliptic example for large, positive $b_0$ so that
$\theta$-dependent contributions to the wall function are small
relative to $b_0$.  The angles $\alpha$ and
$\theta$ are almost equal, and the resulting body is almost a circle.  For
sufficiently large $b_0$, this will hold for all $b( \theta )$.  Then,
\begin{equation}
b_0 + b\left( \alpha + \tan^{-1} \left( \frac{R'}{R} \right) \right) =
\frac{R^2}{\sqrt{R^2 + R'^2}}
\end{equation}
with
\begin{equation}
R = b_0 + r( \alpha ) \ \ .
\end{equation}
It is understood that $b_0 >> r$.  This inequality entitles us to expand
equation (28) in powers of $r/b_0$ and $r'/b_0$ which yields
\begin{eqnarray}
\alpha & = & \theta \\ \nonumber
R( \alpha ) & = & b_0 + b( \alpha ) \ \ .
\end{eqnarray}
Further, by considering equation (28), we see that the curvature of this
shape will always be positive for sufficiently large $b_0$ independent of the
form of $b( \theta )$.

\section{Conclusions}
\label{endup}

We have demonstrated that every two-dimensional rigid body making elastic
collisions between parallel walls is equivalent to a traditional billiard
problem in which a point particle makes specular reflections with walls of
uniquely determined shape.  Certain shapes of particular geometrical interest
may encourage the consideration of specific billiard problems which might
otherwise have escaped attention.  Rectangles and shapes of constant width may
warrant investigation.  Strongly asymmetric rigid bodies strike us as being
of particular interest.  We note that the present results can be extended to
include the effects of external fields.

Similarly, every such billiard problem can be associated with a family of
equivalent rigid body problems.  This offers the possibility of viewing the
extensive billiard literature in a new and potentially revealing light.
This equivalence can be extended.  It is evident, for example, that similar
results can be obtained for a two-dimensional body confined in a rectangular
box.  There is, of course, nothing to preclude the consideration of billiard
problems in higher dimensional spaces.  We believe that the existence of
simple physical systems can give such systems an intuitive immediacy which they
might otherwise lack. Also, the motion of a rigid body in three and
higher dimensions is associated with a non-commutative group.
Thus, we expect it to be equivalent to a new type of billiard
motion.

Finally, the analogous quantum mechanical problem of rigid bodies confined
between parallel walls is of evident interest and is a current focus of our
attention. Here, the features of the quantal results require
a new interpretation, due to the new association with a different
physical property, the tunneling from heads to tails.

We would like to acknowledge helpful discussions with Dr. Serge Troubetzkoy.
One of us (R.C.) would like to acknowledge helpful discussions with
M. Khanji.
This work was supported in part by the US Department of Energy under grant
No.~DE-FG02-88ER 40388 (A.D.J. and R.C.)
and by the National Science Foundation (N.L.B.).

\section{Appendix}
\label{appendix}

Here we shall start from the wall shape obtained for an elliptic disc and
consider the various rigid body shapes which result from different choices of
$b_0$.  Specifically, we take
\begin{equation}
b( \theta ) = b_0 + \left[ \frac{1 + \epsilon \cos^2 \theta }
                                {1 + \epsilon} \right]^{1/2} \ \ .
\end{equation}
In the vicinity of $\alpha = 0$, we find
\begin{equation}
R( \alpha ) = (b_0 + 1) - \frac{1}{2} \ \frac{(1+b_0 ) \epsilon}
{1 + b_0 (1 + \epsilon) } \ \alpha^2 + \ldots \ \ .
\end{equation}
It is useful to recall that the curvature of $R( \alpha )$ is
\begin{equation}
K( \alpha ) = \frac{R^2 + 2 R'^2 - R R''}{(R^2 + R'^2 )^{3/2}}
\end{equation}
where primes indicate derivatives with respect to $\alpha$.  The curvature
corresponding to equation (24) at $\alpha = 0$ is thus
\begin{equation}
K(0) = \frac{1+ \epsilon}{1 + b_0 (1 + \epsilon )} \ \ .
\end{equation}
Evidently, there is a critical value of $b_0$, $b_c = -1/(1+\epsilon )$ such
that $K( 0 )$ is negative for $b_0 < b_c$.  Since the wall function was
assumed to come from the convex hull of the original rigid
body, this result is unphysical.  There is no corresponding
rigid body for $b_0 < b_c$.  It is instructive to describe what happens as one
integrates from $\alpha = \pi / 2$ to $\alpha = 0$
in order to obtain $R( \alpha )$:

{\em Case I. ( $b > b_c$)\/}  Here, $R( \alpha )$ approaches $b_0 + 1$
quadratically (as expected) and $R' ( \alpha )$ approaches
$0$ linearly as $\alpha \rightarrow 0$.  The result is a physical shape
which depends on $b_0$.  The result is consistent in the sense that
the original wall function can be reconstructed from this
$R( \alpha )$ using the general prescription of section II.

{\em Case II. ( $b = b_c$)\/}  This is the limiting physical case.
Now, $R( \alpha )$ approaches $b_0 + 1$ like $\alpha^{4/3}$ and
$R' ( \alpha )$ goes like $\alpha^{1/3}$.  Clearly, $R''$ is badly behaved at
$\alpha = 0$, but this causes no difficulty.  Again, the result is a
completely physical shape which permits reconstruction the original wall
function.  This critical shape is shown in Fig.4.

{\em Case III. ( $b < b_c$)\/} $R( \alpha )$ approaches $b_0 + 1$
{\em linearly\/}, but $R ' ( \alpha )$ does {\em not\/} approach $0$.
Rather, it approaches some non-zero value, $R '_0$, linearly.  The
putative shape has a discontinuous derivative at $\alpha = 0$
(and $\alpha = \pi$).  To see what this means, return to equation (12).
The range $0_+ < \alpha \le \pi / 2$ corresponds to covering the range
$\theta_c \le \theta \le \pi / 2$ with
\begin{equation}
\theta_c = \tan^{-1} \left( \frac{R '_0}{(b_0 + 1 )} \right)  \ \ .
\end{equation}
The finite region between $- \theta_c < \theta < + \theta_c$ is
all governed by the infinitesimal region $0_- < \alpha < 0_+$.  There is
nothing wrong with a single point on the body determining the wall shape
for a finite angular interval.  This occurs whenever the rigid body has a
corner and $R( \alpha )$ has a discontinuous derivative.  However, as in
equation (5), this should correspond to a shape $(b_0 + 1) | \cos \theta |$.
Evidently, this is not the shape of $b_0 + b( \theta )$ chosen here
nor will it be for other choices of $b( \theta )$.  Hence, the resulting
body shape will fail to reproduce $b_0 + b( \theta )$ over the range
$- \theta_c < \theta < \theta_c$, and thus $R( \alpha )$ is not acceptable.
We remark that, in the present example, $b_0 + b( \theta )$
remains positive (and thus not manifestly unphysical) for values of $b_0$ as
small as $-1/\sqrt{1 + \epsilon}$ which is less than the limiting value of
$b_c$ quoted above.

Similar estimates on $b_c$ based on the existence of regions of negative
curvature at the extrema of $b( \theta )$ can be made for arbitrary
(differentiable) wall functions since these extrema necessarily
coincide with the extrema of $R( \alpha )$.  Negative curvature always
indicates an unphysical shape.  While we have not demonstrated
that the first appearance of negative curvature will be at an extremum, these
ease of these estimates suggests that they may be useful.

\begin{figure}
\caption{Phase portrait (in units of $ \pi $) of the contact point
$\theta _c $ versus the angle of incidence $\phi _c $ with respect
to the local normal $\theta _c $ for the elliptic system $\epsilon = 0.03$
and $H = 3$.}
\label{fig1}
\end{figure}
\begin{figure}
\caption{Phase space diagram for the billiard system with $\epsilon = 0.55$
and $H = 1.9$.}
\label{fig2}
\end{figure}
\begin{figure}
\caption{The rigid body shape ($ \vec{R}(\alpha)$)
corresponding to the limiting physical
case of $b_0 = b_c = -1/2$, ($\epsilon = 1$).}
\label{fig3}
\end{figure}

\end{document}